\documentclass[onecolumn,showpacs,amsmath,amssymb,aps,pra,superscriptaddress,floatfix]{revtex4-1}
\usepackage{amsthm}
\usepackage{graphicx}
\usepackage{color}
\usepackage{bm}%
\usepackage{tikz}
\usepackage{subfigure}
\usepackage{caption2}
\usepackage{placeins}
\usepackage{wrapfig}
\usepackage{mathtools}
\usepackage{dsfont}

\begin{document}
\title{Quantum coherence and correlation measures based on affinity}

\author{R. Muthuganesan}
\affiliation{Centre for Nonlinear Science and Engineering, School of Electrical and Electronics Engineering, SASTRA Deemed University, Thanjavur-613 401, Tamil Nadu, India.}

\author{V.K. Chandrasekar}
\affiliation{Centre for Nonlinear Science and Engineering, School of Electrical and Electronics Engineering, SASTRA Deemed University, Thanjavur-613 401, Tamil Nadu, India.}

\author{R. Sankaranarayanan}
\affiliation{Department of Physics, National Institute of Technology, Tiruchirappalli--620015, Tamil Nadu, India}

\begin{abstract}
Coherence and correlation are key features of the quantum system. Quantifying these quantities are astounding task in the framework of resource theory of quantum information processing. In this article, we identify an affinity-based metric to quantify closeness between two states. Using this metric, we introduce a valid quantum coherence measure It is shown that the affinity based coherence measure is bounded by that based on fidelity and trace distance. Further, we propose a bipartite quantum correlation measure based on the affinity metric. The connection between the quantum correlation of states and its local coherence is established. The measure of quantumness in terms of difference of bipartite coherence and corresponding product state coherence is also identified. Finally, we interpret the operational meaning of the affinity based coherence as an upper bound of interferometric power of the quantum state.
\end{abstract}


\maketitle

\section{Introduction}
Quantum coherence and correlation are characteristic features that mark a fundamental departure from the classical realm and notable resources for various information processing \cite{Nielsen}. In quantum information theory, the most interesting and important problems are quantification of resources for various quantum advantageous and classification of states based on the availability of resources.  Coherence plays a central role in emergent fields such as metrology \cite{Giovannetti2004,Demkowicz2014}, biology \cite{Sarovar2010,Lloyd2011,Huelga2013,Lambert2013}, thermodynamics and nanoscale physics \cite{Aberg,Lostaglio}. More recently, application of quantum coherence in refrigeration is also experimentally demonstrated \cite{Buffoni2019}. 

In fact, experimentally realizable coherence measure \cite{Girolami2014coh} and its properties enhance our understanding of coherence \cite{Yu2017}. From the seminal work of Baumgratz et. al. \cite{Baumgratz2014},  various coherence measures proposed based on entropic measures  (relative entropy \cite{Bu2017}, Tsallis  entropy \cite{Zhao2018} and Renyi entropy \cite{Zhu}), distance measures  (trace distance \cite{Rana2016,Chen2018QINP,Wang2016}, Hellinger distance \cite{Jin} and fidelity \cite{Liu2017}) and  information theoretic quantities (quantum Fisher information \cite{Feng2017} and skew information \cite{Girolami2014coh,Yu2017,Baumgratz2014}) are extensively studied. Similar to coherence, quantum correlation is also a consequence of the superposition principle and  is regarded as a useful resource for various information processing tasks. Apart from the entanglement, different measures of quantumness namely quantum discord \cite{Ollivier}, geometric discord \cite{Dakic2010}, measurement-induced nonlocality \cite{Luo2011PRL} have also studied in detail. 

In this article, we introduce a new geometric measure of coherence in terms of  affinity induced metric  which is shown to satisfy all the requirements of a good quantum coherence measure including strong monotonicity condition. Moreover, this quantity  enjoys an elegant analytic expression for pure and single qubit state. It is shown that the affinity based coherence is a lower bound of fidelity based coherence measure and useful quantity in quantum metrology. We present a symmetric  quantum correlation based on the affinity, which can be analytically solvable  for qubit-qudit states. The trade off relation between the quantum coherence and quantum correlation is also derived explicitly. Using the difference of coherence between bipartite state and its product state, we identify a new quantumness measure. The operational meaning of the proposed affinity measure is interpreted in terms of interferrometric power of quantum states.

This paper is organized  as follows: In Sec. \ref{coherence}, we briefly review the concept of quantum coherence. In Sec. \ref{affinity}, we introduce affinity based coherence and establish a relationship with the other coherence measures.  In Sec. \ref{affinitycorre}, we study the symmetric nonclassical correlation based on the affinity distance. Sec. \ref{measure} presents the measure of quantumness interms of coherence. In Sec. \ref{applications}, we explore the possible application of proposed measure in the resource theory. 

\section{Quantum Coherence}
\label{coherence}
In this section, we first review the framework concerning quantifying quantum coherence. Let $\{ |i\rangle, i=0,1, \cdots, d-1 \} $ be  a set of  orthonormal basis with finite dimension $d$. A  state is said to be an incoherent state if the corresponding density operator $\delta$ of the state is diagonal in this basis and denoted as $\delta=\sum_{i=0}^{d-1}\delta_i|i\rangle\langle i|$. We label this set of incoherent quantum states as $\mathcal{I}$: 
\begin{align}
  \mathcal{I}=\left\{\delta =\sum_{i=0}^{d-1}\delta_i|i\rangle\langle i| \right\}.  \label{incohstate}
\end{align}
It is well-known that the incoherent state $\delta$ can have coherence in any other basis. All other states, which are not belonging to the above-mentioned set in the basis, are called coherent states.  In coherence theory, the completely positive and trace-preserving (CPTP) maps which admit an incoherent Kraus representation \cite{Nielsen}. That is, there always exists a set of Kraus operators $\{ K_i\}$ such that 
\begin{align}
\Phi(\delta)\equiv \frac{K_i \delta K_i^{\dagger}}{\text{Tr}(K_i \delta K_i^{\dagger})} \in \mathcal{I}~; ~~~~~~ \sum_i K_iK_i^{\dagger}=\mathds{1}
\end{align}
for all $i$ and any  incoherent state $\delta$. These operations are called incoherent operations.

A functional $C(\rho)$ is a bonafide measure of quantum coherence of a state $\rho$, if it fulfills the following essential requirements \cite{Girolami2014coh,Baumgratz2014}. They are
\begin{enumerate}
\item[(C1)] \textit{Faithfulness}: $C(\rho) \geq 0$ is nonnegative and $C(\rho)=0$ if and only if $\rho$ is an incoherent state.

\item[(C2)] \textit{Monotonicity}: $C(\rho)$ does not increase under the action of an incoherent operation i.e., $C(\rho)\geq C(\Phi(\rho))$.

\item[(C3)] \textit{Strong Monotonicity}: $C(\rho)$ does not increase on average under selective incoherent operations, i.e., $C(\rho)\geq C(\sum_ip_i\rho_i))$, where $p_i=\text{Tr}(K_i \rho K_i^{\dagger})$ and $\rho_i= K_i \rho K_i^{\dagger}/p_i$ with $\sum_i K_i K_i^{\dagger}=\mathds{1}$.

\item[(C4)] \textit{Convexity}: Nonincreasing under mixing of quantum states i.e., $C(\sum_np_n\rho_n)\leq \sum_n p_nC(\rho_n)$ for any set of pure states $\rho_n$ with $\sum_np_n=1$.
\end{enumerate}

It is obvious that conditions (C3) and (C4) imply condition (C2). A quantity that satisfies conditions (C1)--(C3) is called a coherence monotone. If it satisfies condition (C4) in addition, then we call it a convex coherence monotone. In what follows, we identify a new variant of coherence measure based on affinity.

\section{Affinity based Coherence}
\label{affinity}
Affinity, like fidelity characterizes the  closeness of two quantum states. Mathematically, for any two  states $\rho$ and $\sigma$, affinity is defined as \cite{Kholevo1972,Luo2004}
\begin{align}
  \mathcal{A}(\rho,\sigma)=\text{Tr}(\sqrt{\rho}\sqrt{\sigma}).
\end{align}
This definition is much similar to the, Bhattacharya coefficient between two classical probability distribution \cite{Bhattacharyya}. Further, it is worth mentioning that affinity possesses all the properties of  fidelity, which is introduced by Jozsa \cite{Jozsa1994}. Using cyclic property of trace, one can rewrite the affinity as $\mathcal{A}(\rho,\sigma)=\text{Tr}[ (\rho^{1/4}\sigma^{1/4})(\rho^{1/4}\sigma^{1/4})^{\dagger}] $ and we  easily  see  that  the affinity is non-negative. Further, this quantity is more useful in quantum information theory, in particular, entanglement detection, quantum detection, and estimation theory. We  now  list the fundamental  properties  of  the  affinity.
\begin{enumerate}
\item[($\mathcal{A}$1)] $0\leq \mathcal{A}(\rho,\sigma)\leq 1$ and $\mathcal{A}(\rho,\sigma)=1$ if and only if $\rho=\sigma$. Moreover, $\mathcal{A}(\rho,\sigma)=\mathcal{A}(\sigma,\rho)$.

\item[($\mathcal{A}$2)] Affinity is unitary invariant i.e., $\mathcal{A}(\rho,\sigma)=\mathcal{A}(U\rho U^{\dagger},U\sigma U^{\dagger})$ for any unitary operator $U$. 

\item[($\mathcal{A}$3)] Affinity is multiplicative under tensor product:
\begin{align}
  \mathcal{A}(\rho_1 \otimes \rho_2, \sigma_1\otimes \sigma_2)= \mathcal{A}(\rho_1\otimes\sigma_1) \cdot \mathcal{A}(\rho_2\otimes\sigma_2). \nonumber
\end{align}

\item[($\mathcal{A}$4)] $\mathcal{A}(\rho,\sigma)$ is monotonic under CPTP map $\Lambda$ i.e., $\mathcal{A}(\Lambda(\rho),\Lambda(\sigma))\geq\mathcal{A}(\rho,\sigma)$

\item[($\mathcal{A}$5)]  For any orthogonal projectors $\Pi_i=|i\rangle \langle i|$, $\mathcal{A}(\sum_i \Pi_i\rho\Pi_i,\sum_i\Pi_i\sigma\Pi_i)=\sum_i\mathcal{A}(\Pi_i\rho\Pi_i,\Pi_i\sigma\Pi_i)$

where $\rho_i=K_i\rho K_i^{\dagger}/p_i$ and $\sigma _i=K_i \sigma K_i^{\dagger}/q_i$ with $p_i=K_i\rho K_i^{\dagger}$ and $q_i=K_i \sigma K_i^{\dagger}$ are the respective probabilities after the super selection.
\end{enumerate}

To show the property ($\mathcal{A}$4), for any CPTP map, we have to show that $\mathcal{A}(\Lambda(\rho),\Lambda(\sigma))\geqslant  \mathcal{A}(\rho, \sigma)$ where $\Lambda(\sigma)=\sum_i V_i^{\dagger}\sigma V_i$ with $\sum_i V_i^{\dagger} V_i=\mathds{1}$. It is well known that a complete measurement can always be represented as a unitary operation and partial tracing on  an  extended  Hilbert  space $\mathcal{H} \otimes \mathcal{H}_n$ \cite{Vedral}. Let $|j\rangle $ be an orthonormal basis in $\mathcal{H}_n$ and $|\alpha\rangle $ be a unit vector. Then, we can define 
\begin{align}
W=\sum_j V_j \otimes |j\rangle \langle \alpha|
\end{align}
and $W^{\dagger}W=\mathds{1} \otimes |\alpha\rangle \langle \alpha|$. Then there exists a global unitary operator $U$ on $\mathcal{H} \otimes \mathcal{H}_n$. Consequently, 
\begin{align}
U(\rho \otimes|\alpha\rangle \langle \alpha|)U^{\dagger} = \sum_{jj'} V_j \rho V_{j'}^{\dagger} \otimes |j\rangle \langle j'|   \nonumber
\end{align}
and 
\begin{align}
\text{Tr}_{\mathcal{H}_n}(U(\rho \otimes|\alpha\rangle \langle \alpha|)U^{\dagger} )=\sum_{j} V_j \rho V_{j}^{\dagger}=\Lambda(\rho).
\end{align}
Assume $\rho^j$ and $\sigma^j$ are the marginal states of $\rho$ and $\sigma$ respectively. Then $\mathcal{A}(\rho^j, \sigma^j)\geq \mathcal{A}(\rho, \sigma)$ \cite{Luo2004},  using  unitary invariant ($\mathcal{A}$2), and multiplicativity ($\mathcal{A}$3), one can write
\begin{align}
\mathcal{A}(\Lambda(\rho),\Lambda(\sigma))=&\mathcal{A}(\text{Tr}_{\mathcal{H}_n}(U(\rho\otimes|\alpha\rangle \langle \alpha|)U^{\dagger} ),\text{Tr}_{\mathcal{H}_n}(U(\sigma\otimes|\alpha\rangle \langle \alpha|)U^{\dagger} ))\geq \mathcal{A}(\rho, \sigma).\large 
  \label{property4}
\end{align}

Next, we provide proof of property ($\mathcal{A}$5). Let $\{ \Pi_i\} $ be a set of orthogonal projectors satisfy the conditions $\sum_i\Pi_i=\mathds{1}$ and $\Pi_i\Pi_j=\Pi_i \delta_{ij}$. We have
\begin{align}
\mathcal{A}\left(\sum_i\Pi_i\rho\Pi_i,\sum_i\Pi_i\sigma\Pi_i\right)=\text{Tr}\sqrt{\sum_i\Pi_i\rho\Pi_i}\sqrt{\sum_j \Pi_j\sigma\Pi_j}, \nonumber
\end{align}
\begin{align}
=\text{Tr}\sum_i\sqrt{\Pi_i\rho\Pi_i}\sum_j\sqrt{\Pi_j\sigma\Pi_j} =\sum_i\text{Tr}\sqrt{\Pi_i\rho\Pi_i}\sqrt{\Pi_i\sigma\Pi_i},  \label{Monotonicity}
\end{align}
\begin{align}
=\sum_i\mathcal{A}\left(\Pi_i\rho\Pi_i, \Pi_i\sigma\Pi_i \right).    \nonumber
\end{align}

Though affinity itself is not a metric, due to monotonicity and concavity property of affinity \cite{Luo2004}, one can define  any monotonically decreasing function of affinity  as a metric in state space. One such metric in terms of affinity is defined as 
\begin{align}
  d_{\mathcal{A}}(\rho,\sigma)=1-\mathcal{A}(\rho,\sigma).
\end{align}
Recently, the usefulness of the metric in characterizing nonclassical correlation is demonstrated  \cite{Muthuaffinity}. Based on the above metric, we propose a new version of quantum coherence monotone. Defining quantum coherence in terms of affinity as 
\begin{align}
C_{\mathcal{A}}(\rho)=~^\text{min}_{\delta \in \mathcal{I}}~ d_{\mathcal{A}}(\rho,\delta)=1-~^\text{max}_{\delta \in \mathcal{I}}~ {\mathcal{A}}(\rho, \delta),
\end{align} 
where the optimization is taken over all possible incoherent state from the set as given in Eq. (\ref{incohstate}). Using the properties of affinity ($\mathcal{A}$1)--($\mathcal{A}$6), affinity distance is a valid coherence measure and coherence monotone.  Since $\rho=\sum_n p_n|\phi_n\rangle\langle \phi_n|$, we can also write affinity-based quantum coherence via convex--roof construction as 
\begin{align}
C_{\mathcal{A}}(\rho)=~~^\text{min}_{\{ p_n,|\phi_n\rangle \} }\sum_n p_n C_{\mathcal{A}}(|\phi_n\rangle ),
\end{align} 
where the minimum is taken over all the ensembles $\{ p_n,|\phi_n\rangle \} $. 

\subsection{Bounds on Coherence}
Defining coherence measure based on fidelity as \cite{Liu2017}
\begin{align}
C_{\mathcal{F}}(\rho)=1-~^\text{max}_{\delta \in \mathcal{I}} ~\mathcal{F}(\rho, \delta),
\end{align} 
where $\mathcal{F}(\rho, \delta)=\text{Tr}\sqrt{\rho^{1/2} \delta \rho^{1/2}}$ is the fidelity between the states $\rho$ and $\delta$. Alternatively, the fidelity can be expressed as 
\begin{align}
\mathcal{F}(\rho, \delta)=\text{Tr}\left( U \sqrt{\rho} \sqrt{\delta}\right), 
\end{align} 
where $U$ is a unitary operator. We conclude that $\mathcal{F}(\rho, \delta)\geq  \mathcal{A}(\rho, \delta)$, then
\begin{align}
C_{\mathcal{F}}(\rho)\leq C_{\mathcal{A}}(\rho).  \label{lower}
\end{align} 
In other words, the fidelity based coherence measure is always bounded by the affinity based coherence. We emphasize that affinity can be employed in an equivalent way with another figure of merit in discriminating between two quantum states, namely, their trace distance. This property was found long ago by Holevo, who proved the following pair of inequalities \cite{Kholevo1972}
\begin{align}
1-\mathcal{A}(\rho,\sigma)\leq T(\rho,\sigma)\leq1-[\mathcal{A}(\rho,\sigma)]^2,
\end{align} 
where $T(\rho,\sigma)$ is the trace distance between the states $\rho$ and $\sigma$. The trace metric is  particularly important measure owing to its connection with the probability of error between two quantum states and characterizing the nonclassicality.  One can define coherence measure based on the trace distance as \cite{Rana2016,Chen2018QINP,Wang2016}
\begin{align}
C_T(\rho)=~~^\text{min}_{\delta \in \mathcal{I} }\lVert \rho- \delta\rVert_1. 
\end{align} 
Then, it is observed that the upper bound of affinity-based coherence is
\begin{align}
C_{\mathcal{A}}(\rho)\leq C_T(\rho). \label{upper}
\end{align} 
Combining Eqs. (\ref{lower}) and (\ref{upper}), we obtain the bound for affinity of coherence
\begin{align}
C_{\mathcal{F}}(\rho)\leq C_{\mathcal{A}}(\rho)\leq C_T(\rho)
\end{align}
implying that the fidelity and trace distance based coherence measures are lower and upper bounds of affinity based coherence measures respectively. 

\subsection{Alternative Expression}
Let $\delta=\sum_i \delta_i |i\rangle \langle i|$ be an incoherent state. Then the affinity between the state of our interest and incoherent state is written as 
\begin{align}
\mathcal{A}(\rho, \delta)= \text{Tr}\left(\sqrt{\rho} \sqrt{\delta} \right)=\sum_i \sqrt{\delta_i}\langle i|\sqrt{\rho}|i\rangle. \nonumber
\end{align} 
According to Cauchy-Schwarz inequality, we have \cite{Yu2017}
\begin{align}
\left(\sum_{i=0}^{d-1} \frac{\sqrt{\delta_i} \langle i|\sqrt{\rho}|i\rangle}{M}\right)^2\leq \left(\sum_{i=0}^{d-1} \frac{\sqrt{\delta_i} \langle i|\sqrt{\rho}|i\rangle^2}{M^2}\right) \left(\sum_{i=0}^{d-1}\delta_i \right)=1.
\end{align} 
The above inequality saturates when 
\begin{align}
\sqrt{\delta_i}= \frac{\langle i|\sqrt{\rho}|i\rangle}{M}      \nonumber
\end{align}
and we obtain 
\begin{align}
^\text{max}_{\delta \in \mathcal{I}} \mathcal{A}(\rho, \delta)= M= \sqrt{\sum_{i=0}^{d-1}\langle i|\sqrt{\rho}|i\rangle^2}.
\end{align}
Then the affinity of coherence is computed as 
\begin{align}
C_{\mathcal{A}}(\rho)=1-\sqrt{\sum_{i=0}^{d-1}\langle i|\sqrt{\rho}|i\rangle^2} \nonumber
\end{align}
and the closest incoherent (optimal) state is 
\begin{align}
\delta=\delta_0=\sum_i\frac{\langle i|\sqrt{\rho}|i\rangle^2}{\langle i'|\sqrt{\rho}|i'\rangle^2} |i\rangle \langle i|. \nonumber
\end{align}
For pure state $\sqrt{\rho}=\rho$, the affinity-based coherence is 
\begin{align}
C_{\mathcal{A}}(\rho)=1-\sqrt{\sum_{i=0}^{d-1}\langle i|\rho|i\rangle^2}.
\end{align}

\subsection{Coherence of single qubit state}

 By considering a Bloch sphere representation, quantum states $\rho$ and $\delta$ of single qubit can be expressed as
\begin{align}
\rho = \frac{\mathds{1}_2+\mathbf{r}.\sigma}{2}, ~~~~~~~~~~~~\delta = \frac{\mathds{1}_2+\mathbf{s}.\sigma}{2},
\end{align} 
where $\mathds{1}_2$ is $2 \times 2$ identity matrix, $\mathbf{r}$ and $\mathbf{s}$ are Bloch vectors and $\sigma$ is a vector of Pauli spin matrices. The affinity between the states, $\rho$ and $\delta$ read as \cite{Luo2004}
\begin{align}
\mathcal{A}(\rho, \delta)= \frac{(1+\sqrt{1-\lvert \mathbf{r} \rvert^2})(1+\sqrt{1-\lvert \mathbf{s} \rvert^2})+\mathbf{r}\cdot\mathbf{s}}{(\sqrt{ 1+\lvert\mathbf{r} \rvert}+ \sqrt{1-\lvert\mathbf{r} \rvert})(\sqrt{ 1+\lvert\mathbf{s} \rvert}+ \sqrt{1-\lvert\mathbf{s} \rvert})},
\end{align} 
where $\mathbf{r}\cdot\mathbf{s}=r_x s_x+r_y s_y+r_z s_z$ is the dot product,  $\lvert\mathbf{r} \rvert$ and $\lvert \mathbf{s} \rvert$ are the magnitudes of vector $\mathbf{r}$ and $\mathbf{s}$ respectively.  Since $\delta$ is an incoherent state, then the vector  $\mathbf{s}$ can be represented as $\mathbf{s}=(0,0,s_z)$. For pure state $\lvert\mathbf{r} \rvert=1$,  the affinity can be recast as 
\begin{align}
\mathcal{A}(\rho, \delta)= \frac{1+\sqrt{1-s_z^2}+r_zs_z}{\sqrt{2} \left(\sqrt{ 1+s_z}+ \sqrt{1-s_z}\right)}.
\end{align} 
In order to get the optimal value of affinity of coherence, the maximization is taken over $s_z$. 

\section{Quantum Correlation}
\label{affinitycorre}
It is well known fact that entanglement and coherence are interrelated via incoherent operations. In order to establish a relation between the measures of  coherence and quantum correlation, we introduce quantum correlation measure based on affinity. The symmetrized  affinity-based  geometric discord is defined as 
\begin{align}
D_{\mathcal{A}}(\rho)=~^\text{min}_{\Pi^{ab}}~d_{\mathcal{A}}\left(\rho, \Pi^{ab}(\rho)\right)=1-~~^\text{max}_{\Pi^{ab}}~\mathcal{A}\left( \rho, \Pi^{ab}(\rho)\right), \label{Affinitycorr}
\end{align}
where the maximum is taken over all possible locally invariant projective measurements on subsystems $a$ and $b$. Here, $\Pi^{ab}(\rho)=\sum_{k, k'} (\Pi^a_k \otimes \Pi^b_{k'}) \rho (\Pi^a_{k} \otimes \Pi^b_{k'})$ is the post-measurement state with $\Pi^a=\{\Pi_k^a= |k\rangle \langle k|\} $ and $\Pi^b=\{\Pi_{k'}^b= |k'\rangle \langle k'|\} $, which do not change the marginal states i.e.,
\begin{align}
\rho^a=\sum_k \Pi_k^a \rho^a \Pi_k^a  ~~~~~~~ \text{and} ~~~~~~~ \rho^b=\sum_{k'} \Pi_{k'}^b \rho^b \Pi_{k'}^b.                \nonumber
\end{align}

The quantity $D_{\mathcal{A}}(\rho)$ satisfies all necessary axioms of a bonafide measure of quantum correlation measure. Here, we demonstrate some interesting properties of affinity-based quantum correlation measure. 
\begin{enumerate}
\item[(i)]  $D_{\mathcal{A}}(\rho)$ is non-negative i.e., $D_{\mathcal{A}}(\rho)\geq 0$. 

\item[(ii)] $D_{\mathcal{A}}(\rho)=0$ for any product state $\rho=\rho_{a}\otimes  \rho _{b}$ and the classical state in the form $\rho =\sum _{k,k'}p_{kk'}|k\rangle \langle k| \otimes |k'\rangle \langle k'|$. 

For any product and classical state, one can always find $\Pi_k^{ab}$ such that $\rho=\Pi^{ab}(\rho)$  and $\mathcal{A}(\rho,\Pi^{ab}(\rho))=1$, which leads to zero discord.

\item[(iii)] $D_{\mathcal{A}}(\rho)$ is locally unitary  invariant i.e., $D_{\mathcal{A}}\left((U\otimes   V)\rho  (U\otimes   V)^\dagger\right)=D_{\mathcal{A}}(\rho)$ for any local unitary operators $U$ and $V$. 

\textit{Proof}: Let $\rho'=(U\otimes V)\rho  (U\otimes V)^\dagger$, we have
\begin{align}
D_{\mathcal{A}}\left(\rho'\right)=& ~^{\text{min}}_{\Pi^{ab}}~d_{\mathcal{A}}\left(\rho', \Pi^{ab}(\rho')\right) \nonumber \\
=& ~^{~~~\text{min}}_{\{ \Pi_k^{a}, \Pi_{k'}^{b}\} } \sum_{k, k'}d_{\mathcal{A}}\left(\rho, (U\Pi_k^aU^{\dagger}) \otimes (V\Pi_{k'}^bV^{\dagger}) \rho (U\Pi_k^aU^{\dagger}) \otimes (V\Pi_{k'}^bV^{\dagger}) \right)  \nonumber\\
=& ~^{~~~\text{min}}_{\{ \Pi_k^{A}, \Pi_k^{B}\} } \sum_{k, k'} d_{\mathcal{A}}\left(\rho,  (P^a_k \otimes P^b_{k'}) \rho (P^a_k \otimes P^b_{k'}) \right) \nonumber \\
=& D_{\mathcal{A}}(\rho).    \nonumber
\end{align}
Here $P^a_k=(U\Pi_k^aU^{\dagger})$ and $P^b_{k'}=(V\Pi_{k'}^bV^{\dagger})$ are also the eigen-projectors of the marginal states $\rho^a$ and $\rho^b$ respectively.

\item[(iv)] For any arbitrary $m \times n$ dimensional pure state with $m\leq n$, the affinity-based discord is $D_{\mathcal{A}}(\rho) = 1-\sum_i s_i^2$, where $s_i$ are Schmidt coefficients. 

\textit{Proof}: For an arbitrary $m \times n$ dimensional pure state with $m\leq n$, the Schmidt decomposition is given as 
\begin{align}
|\Psi\rangle =\sum_i \sqrt{s_i} |\alpha_i \rangle \otimes |\beta_i\rangle,  \nonumber
\end{align}
where $s_i$ is Schmidt's coefficient, $|\alpha_i \rangle$ and $|\beta_i\rangle$  are orthonormal bases of subsystems $\rho^a$ and $\rho^b$ respectively. Using the identity $\Pi^{ab} f(\rho)\Pi^{ab}=f(\Pi^{ab} \rho \Pi^{ab})$ \cite{Girolami2012}, one can rewrite the discord as 
\begin{align}
D_{\mathcal{A}}(\rho)=~1-~^{~~\text{max}}_{\{ \Pi_k^{a}, \Pi_{k'}^{b}\} } \sum_k\text{Tr}[ \sqrt{\rho} (\Pi_k^{a}\otimes \Pi_{k'}^{b})\sqrt{\rho}(\Pi_k^{a}\otimes \Pi_{k'}^{b})]. \label{identity}
\end{align}
After a straight forward calculation the affinity between pre- and post-measurement state is $\sum_i s_i^2$. Then the affinity-based geometric discord is computed as 
\begin{align}
D_{\mathcal{A}}(\rho)=1-\sum_i s_i^2,
\end{align}
which is equal to the geometric measure of entanglement and other few measures. The quantity  $\sum_i s_i^2$ is bounded by $1/m$. Hence, $D_{\mathcal{A}}(\rho)\geq (m-1)/m$ and the inequality saturates for pure maximally entangled state.  

\item[(v)] For any arbitrary $m \times n$ dimensional mixed state, the affinity--based discord has a lower bound as 
\begin{align}
D_{\mathcal{A}}\left(\rho\right)\geq 1-\sum_{i=1}^{\text{min}\{ m-1,n-1\} }\mu_i=\text{max}\{ D^a_{\mathcal{A}}\left(\rho\right),D^b_{\mathcal{A}}\left(\rho\right)\},
\end{align}
where $D^{a(b)}_{\mathcal{A}}\left(\rho\right)$ is affinity-based correlation measure while measuring on $a(b)$.

\textit{Proof}:An arbitrary state $\rho$ of a bipartite composite system can be written as 
\begin{align}
\sqrt{\rho}=\sum_{i,j} \gamma_{ij} X_i \otimes Y_j,
\end{align}
where $\gamma_{ij}=\text{Tr}(\sqrt{\rho} X_i \otimes Y_j)$ real elements of correlation matrix $\Gamma$ of order $m^2 \times n^2$, $X_i$  and $Y_j$ are orthonormal self-adjoint observable of state space of marginal states  satisfying the condition $\text{Tr}(X_k X_l)=\text{Tr}(Y_k Y_l)=\delta_{kl}$. After a straight forward calculation, the affinity between the state of our interest and post--measurement state is computed as 
\begin{align}
\mathcal{A}(\rho, \Pi^{ab}(\rho))=~^\text{{min}}_{A,B}~\text{Tr}(A\Gamma B^t B\Gamma^t A^t),    \nonumber
\end{align}
where $A=(a_{ki}=\text{Tr}|k\rangle \langle k |X_i)$ and $B=(b_{k'j}=\text{Tr}|k'\rangle \langle k'|Y_j)$ and $t$ denotes the transpose of a matrix. The affinity-based quantum correlation measure is 
\begin{align}
D_{\mathcal{A}}(\rho)=1-~^\text{{max}}_{A,B}~\text{Tr}(A\Gamma B^t B\Gamma^t A^t). \label{Final}
\end{align}
Following the standard optimization procedure, we have a lower bound for $D_{\mathcal{A}}(\rho)$ due to the projective measurements on $a$ as
\begin{align}
D_{\mathcal{A}}^a(\rho)\geq 1-\sum_{i=1}^{m-1} \mu_i,
\end{align} 
where $\mu_i$ are eigenvalues of matrix $\Gamma \Gamma^t$ arranged in ascending order and due to the projective measurements on $b$, we have the following lower bound as 
\begin{align}
D_{\mathcal{A}}^b(\rho)\geq 1-\sum_{i=1}^{n-1} \mu_i.
\end{align}
From the above inequalities, the affinity-based quantum correlation measure has a tight lower bound as
\begin{align}
D_{\mathcal{A}}(\rho)\geq 1-\sum_{i=1}^{\text{min}\{m-1, ~n-1\} } \mu_i= \text{max}\{D_{\mathcal{A}}^a,~ D_{\mathcal{A}}^b \}
\end{align}
Further, the closed formula of $D_{\mathcal{A}}(\rho)$ for any $2 \times n$ dimensional system is 
\begin{align}
D_{\mathcal{A}}(\rho)=1-\mu_1.   \nonumber
\end{align}

\item[(vi)] The affinity based quantum correlation measure and coherence are related as 
\begin{align}
D_{\mathcal{A}}(\Phi[\rho^a \otimes \rho^b]) \leq  C_{\mathcal{A}} (\Phi[\rho^a \otimes \rho^b]). 
\end{align}
It is well known fact that incoherent operation do not increase the coherence of a state. Let $\Phi$ be an incoherent operation on the product state $\rho=\rho^a \otimes \rho^b$. The affinity-based symmetric quantum correlation measure defined in Eq. (\ref{Affinitycorr}) is equal to the affinity of coherence relative to Luder's measurements as a simple extension of coherence measure relative to von Neumann projective measurements. Using monotonicity, we can write as \cite{Yu2017}
\begin{align}
C_{\mathcal{A}} (\Phi[\rho^a \otimes \rho^b])\leq C_{\mathcal{A}} (\rho^a \otimes \rho^b)=2 \left\{ 1-\left[1-\frac{1}{2}C_{\mathcal{A}}(\rho^a)\right] \left[1-\frac{1}{2}C_{\mathcal{A}}(\rho^a)\right] \right \} \label{conversion} 
\end{align}
and hence 
\begin{align}
D_{\mathcal{A}}(\Phi[\rho^a \otimes \rho^b]) \leq C_{\mathcal{A}} (\rho^a \otimes \rho^b). \nonumber
\end{align}
The equation characterizes the transformation of the local coherence to global quantum correlation under incoherent operations. It is shown that the equality holds while choosing the incoherent operation as \cite{Yu2017}, 
\begin{align}
\Phi=\sum_{ij} |i,i\oplus j\rangle \langle i,j |.
\end{align}
If one of the reduced states $\rho^{a(b)}$ is incoherent, Eq. (\ref{conversion}) becomes
\begin{align}
D_{\mathcal{A}}(\Phi[\rho^a \otimes \rho^b]) \leq C_{\mathcal{A}}(\rho^{b(a)}).
\end{align}
\end{enumerate}

\section{Quantum Coherence relative to Measurements} 
\label{measure}
Consider a bipartite quantum system in the composite separable Hilbert space $\mathcal{H}=\mathcal{H}^a\otimes\mathcal{H}^b$ with respective Hilbert spaces of the marginal systems $a$ and $b$.  Let $\rho$ be any quantum state in  $\mathcal{H}$ shared by parties $a$ and $b$ and the projective measurements $\Pi=\{\Pi_i\} $, i.e., $\Pi_i$ are orthogonal projectors satisfying the conditions $\sum_i\Pi_i=\mathds{1}$ and $\Pi_i\Pi_j=\delta_{ij}\Pi_i$. By generalizing the above-defined coherence measure relative to von Neumann measurements, one can define measurement-based coherence as 
\begin{align}
C_{\mathcal{A}}(\rho|\Pi)=1-~\mathcal{A}(\rho, \Pi(\rho)),
\end{align}
where $\Pi(\rho)=\sum_i(\Pi_i^a\otimes\mathds{1})\rho(\Pi_i^a\otimes\mathds{1})$ is the post-measurement state after the projective measurement on the subsystem $a$. Similarly, the Hellinger-distance based coherence measure is defined as 
\begin{align}
C_{\mathcal{H}}(\rho|\Pi)=\lVert\sqrt{\rho} - \Pi(\sqrt{\rho}) \rVert^2_2,
\end{align}
where $\lVert A \rVert^2_2=\text{Tr}(AA^{\dagger})$. Firstly, we establish a simple relation between affinity of coherence and Hellinger distance based coherence measure. 

{\bf Theorem.1}: \textit{The affinity-based coherence is equal to Hellinger distance-based coherence measure i.e.,
\begin{align}
C_{\mathcal{A}}(\rho|\Pi)=C_{\mathcal{H}}(\rho|\Pi). \nonumber
\end{align}
}
To establish this relation, using the identity  $\Pi^a f(\rho)\Pi^a=f(\Pi^a \rho \Pi^a)$ \cite{{Girolami2012}}, one can rewrite the definition of affinity based coherence measure as 
\begin{align}
C_{\mathcal{A}}(\rho|\Pi)=1- ~\mathcal{A}(\rho, \Pi(\rho))=1- \text{Tr}\left( \sqrt{\rho} \Pi(\sqrt{\rho})\right).
\end{align}
Then, 
\begin{align}
C_{\mathcal{H}}(\rho|\Pi)=\lVert\sqrt{\rho} - \Pi(\sqrt{\rho}) \rVert^2=1-2\text{Tr}\left( \sqrt{\rho} \Pi(\sqrt{\rho})\right)+\text{Tr}\left(\Pi(\sqrt{\rho})\right)^2. \nonumber
\end{align}
Using the orthogonality of projectors and cyclic property of trace, we can show that 
\begin{align}
\text{Tr}\left(\Pi(\sqrt{\rho})\right)^2=\text{Tr}\left( \sqrt{\rho} \Pi(\sqrt{\rho})\right) \nonumber
\end{align}
and thus
\begin{align}
C_{\mathcal{H}}(\rho|\Pi)=1-\text{Tr}\left( \sqrt{\rho} \Pi(\sqrt{\rho})\right)=C_{\mathcal{A}}(\rho|\Pi). \nonumber
\end{align}

Next, we define a quantum correlation measure using the coherence measure based on the affinity. The difference between global coherence and product state coherence is defined as 
\begin{align}
\Delta_{\mathcal{A}}(\rho|\Pi)=C_{\mathcal{A}}(\rho|\Pi)-C_{\mathcal{A}}(\rho^{a}\otimes\rho^{b}|\Pi).   \nonumber
\end{align}
Using the tensor product identity $(A\otimes B)^{1/2}=A^{1/2}\otimes B^{1/2}$ and trace property, we can easily show that $C_{\mathcal{A}}(\rho^{a}\otimes\rho^{b}|\Pi)=C_{\mathcal{A}}(\rho^{a}|\Pi^a)$. Then, the coherence difference can be written as 
\begin{align}
\Delta_{\mathcal{A}}(\rho|\Pi)=C_{\mathcal{A}}(\rho|\Pi)-C_{\mathcal{A}}(\rho^{a}|\Pi^a)
\end{align}
to quantify the coherence difference between global state $\rho^{ab}$ coherence and local state $\rho^a$ coherence. Here, $\Pi^a=\sum_k\Pi^a_k\rho^a\Pi^a_k$ is the projective measurements on subsystem $a$. The quantity $\Delta_{\mathcal{A}}(\rho|\Pi)$ is non-negative and zero if $\rho$ is not perturbed by the measurement i.e., $\rho=\Pi(\rho)$. Defining the correlated coherence in terms of coherence difference is defined as 
\begin{align}
Q_{\mathcal{A}}(\rho)=~^\text{min}_{~\Pi}~\Delta_{\mathcal{A}}(\rho|\Pi),
\end{align}
where the minimization is taken over von Neumann projective measurements. Here, we demonstrate some important properties of the correlated coherence measure as below:
\begin{enumerate}
\item[(i)] For any bipartite quantum state $Q_{\mathcal{A}}(\rho)\geq 0$. The equality holds if and only if the quantum state $\rho$  has the forms $\rho=\sum_{i}p_{i}|i\rangle \langle i|\otimes\sigma_i$. 

\item[(ii)]For any entangled state, $Q_{\mathcal{A}}(\rho)$ is a positive quantity.

\item[(iii)] $Q_{\mathcal{A}}(\rho)$ is invariant under local unitary transformation i.e., $Q_{\mathcal{A}}(\rho)=Q_{\mathcal{A}}(U\rho U^{\dagger})$, where $U=U^a\otimes U^b$ with $U^a$ and  $U^b$  are the unitary operators on marginal Hilbert spaces of $a$ and $b$ respectively.
\item[(iv)] $Q_{\mathcal{A}}(\rho)$ is non-increasing under any channel $\Phi^b$ on subsystem $b$, i.e., $Q_{\mathcal{A}}((\mathds{1}^a \otimes \Phi^{b})(\rho))\leq Q_{\mathcal{A}}(\rho^{ab})$
\end{enumerate}
Proof: (i) It is easy to show the nonnegativity of $Q_{\mathcal{A}}(\rho)$, and so we consider the second part of inequality. For $\rho=\sum_{i}p_{i}|i\rangle \langle i|\otimes\sigma_i$, the subsystem $\rho^a=\sum_{i}p_{i}|i\rangle \langle i|$ is an incoherent state and $\rho$ is unaffected due to von Neumann measurements. Then $C_{\mathcal{A}}(\rho^{a})=C_{\mathcal{A}}(\rho|\Pi)=0$ and leads to zero correlation. 

(ii) To establish property (ii), one can show that both product and separable states are not disturbed by the measurements. Using the above property it is clear that any entangled states are disturbed due to local measurements. The affinity between the pre- and post-measurement states are nonzero and the correlated coherence is positive i.e., $Q_{\mathcal{A}}(\rho)> 0$.

(iii) Let $U$ and $V$ are unitary operators acting on the marginal Hilbert spaces. Then 
\begin{align}
C_{\mathcal{A}}\left((U\otimes V)\rho(U\otimes V)^{\dagger}| \Pi\right)=\sum_k d_{\mathcal{A}}\big[ \left((U\otimes V)\rho  (U\otimes V)^\dagger\right),
&(\Pi_k^{a}\otimes \mathds{1}) \left((U\otimes V)\rho  (U\otimes V)^\dagger\right) (\Pi_k^{a}\otimes \mathds{1})\big],  \nonumber \\
=\sum_k d_{\mathcal{A}}\big[\rho, (U\Pi_k^{a}U^\dagger\otimes\mathds{1})\rho(U\Pi_k^{a}U^\dagger\otimes\mathds{1})\big]  \nonumber\\
=C_{\mathcal{A}}(\rho^{ab}|\Pi).~~~~~~~~~~~~~~~~~~~~~~~~~~~~~~~~~~~~~~~~~   \nonumber
\end{align}
Similarly, one can show that $C_{\mathcal{A}}\left((U\otimes V)\rho^{a}(U\otimes V)^{\dagger}| \Pi^a\right)=C_{\mathcal{A}}(\rho^{a}|\Pi^a)$.

(iv) Let $\Phi^b$ be a quantum operation on subsystem $b$. Then the global coherence is 
\begin{align}
C_{\mathcal{A}}\left((\mathds{1}^a \otimes\Phi^{b})(\rho|\Pi)\right)=d_{\mathcal{A}}\left((\mathds{1}^a \otimes\Phi^{b})(\rho|\Pi)\right)=d_{\mathcal{A}}\left( \text{Tr}_c(\mathds{1}^a\otimes U)(\rho\otimes\rho^{c})(\mathds{1}^a\otimes U)^{\dagger},\text{Tr}_c(\Pi^a\otimes\mathds{1}^b\otimes\mathds{1}^c) \right). \nonumber
\end{align}
Due to the contractivity of affinity distance, we can write 
\begin{align}
C_{\mathcal{A}}((\mathds{1}^a \otimes\Phi^{b})(\rho)|\Pi)\leq d_{\mathcal{A}}\left(\rho\otimes\rho^{c},\Pi^a\otimes\mathds{1}^b\otimes\mathds{1}^c \right)= C_{\mathcal{A}}(\rho|\Pi). \nonumber
\end{align}

Next, we choose the projectors as eigenvectors of the marginal state $\rho^a$, then the marginal states are invariant under these measurements i.e., $\sum_k \Pi_k^a\rho^a\Pi_k^a=\rho^a$. In this situation, $C_{\mathcal{A}}(\rho^{a}|\Pi^a)=0$ and the difference between the global coherence and local coherence becomes 
\begin{align}
\Delta_{\mathcal{A}}(\rho|\Pi)=C_{\mathcal{A}}(\rho|\Pi). \label{diff}
\end{align}
It is worth mentioning that this quantity also synthesizes the amount of quantum correlation contained in the quantum state as 
\begin{align}
N_{\mathcal{A}}(\rho)=~^\text{max}_{~\Pi}~\Delta_{\mathcal{A}}(\rho|\Pi),
\end{align}
where the maximum is taken over the von Neumann projective measurements. This quantity is analog of measurement-induced nonlocality \cite{Luo2011PRL}. Consequently, we have the following relation between the correlation measures and coherence value as
\begin{align}
C^{\text{min}}_{\mathcal{A}}(\rho|\Pi)\geq Q_{\mathcal{A}}(\rho),~~~~\text{and}~~~~~C^{\text{max}}_{\mathcal{A}}(\rho|\Pi)\geq N_{\mathcal{A}}(\rho).
\end{align}

\subsection{Examples}
To gain  more intuitive understanding of the correlated coherence based on the affinity, we now evaluate explicitly the measures of correlations for several typical states.

\textit{Bell diagonal state}:
First, we consider Bell diagonal state and the Bloch representation of the state can be expressed as 
\begin{equation}
\rho^{BD}=\frac{1}{4}\left[\mathds{1}\otimes\mathds{1}+\sum^3_{i=1}c_i(\sigma^i \otimes \sigma^i)\right],
\end{equation}
where the vector $\vec{c}=(c_1,c_2,c_3) $ is a three dimensional vector composed of correlation coefficients such that $-1\leq c_i=\text{Tr}(\rho^{BD}\sigma^i \otimes\sigma^i)\leq 1$ completely specify the quantum state. This class of states are indeed nothing but the convex combination of Bell states given as 
\begin{equation}
\rho^{BD}=\lambda^+_{\phi} |\phi^+\rangle \langle \phi^+|+\lambda^-_{\phi} |\phi^-\rangle \langle \phi^-|+\lambda^+_{\psi} |\psi^+\rangle \langle \psi^+|+\lambda^-_{\psi} |\psi^-\rangle \langle \psi^-|,
\end{equation}
where the non-negative eigenvalues of the density matrix $\rho^{BD}$ read as
\begin{align}
\lambda^{\pm }_{\psi}=\frac{1}{4}[1\pm c_1\mp c_2+c_3], ~~~~~~~~~~ \lambda^{\pm }_{\phi}=\frac{1}{4}[1\pm c_1\pm c_2+c_3] \nonumber
\end{align}
with $|\psi^{\pm}\rangle=[|00\rangle +|11\rangle]/\sqrt{2} $ and $|\phi^{\pm}\rangle=[|01\rangle +|10\rangle]/\sqrt{2} $ are the four maximally entangled states. The reduced state  $\rho^{a(b)}=\text{Tr}_{(a)b}\rho^{BD}=\mathds{1}^{b(a)}/2$ is maximally incoherent state and $C_{\mathcal{A}}(\rho^a|\Pi^a)=0$. The coherence and correlated coherence are computed as
\begin{align}
C^{\text{min}}_{\mathcal{A}}(\rho^{BD}|\Pi)=Q_{\mathcal{A}}(\rho^{BD}|\Pi)=\frac{1}{2}-\left(\sqrt{\lambda^+_{\phi}}\sqrt{\lambda^-_{\phi}}+\sqrt{\lambda^+_{\psi}}\sqrt{\lambda^-_{\psi}}\right). \nonumber 
\end{align}
\textit{Werner state}:
Next, we consider $m\times m-$ dimensional Werner state, which is defined as 
\begin{align}
\omega=\frac{m-\text{x}}{m^3-m}\mathds{1}+\frac{m\text{x}-1}{m^3-m}F,  ~~~~~~~~~~~~\text{x}\in[-1, 1],
\end{align}
with $F=\sum_{kl}|kl\rangle \langle kl|$. Here also we shall note that the marginal states are maximally incoherent state. After a straight forward calculation, we obtained the measures as   
\begin{align}
C^{\text{min}}_{\mathcal{A}}(\omega|\Pi)=Q_{\mathcal{A}}(\omega|\Pi)=\frac{1}{2} \left(\frac{m-\text{x}}{m+1} -\sqrt{\frac{m-1}{m+1}(1-\text{x}^2)}\right).
\end{align}
\textit{Isotropic state}: $m\times m-$ dimensional isotropic state is defined as 
\begin{align}
\rho=\frac{1-\text{x}}{m^2-1}\mathds{1}+\frac{m^2\text{x}}{m^2-1}|\Psi^+\rangle \langle \Psi^+|, ~~~~~~~\text{x}\in[0,1],
\end{align}
where $|\Psi^+\rangle=\frac{1}{\sqrt{m}}\sum_i |ii\rangle $. The affinity-based correlated coherence measure is computed as  
\begin{align}
C^{\text{min}}_{\mathcal{A}}(\rho|\Pi)=Q_{\mathcal{A}}(\rho|\Pi)=\frac{1}{m} \left(\sqrt{(m-1)\text{x}}-\sqrt{\frac{1-\text{x}}{m+1}}\right)^2. \nonumber
\end{align}
\textit{Pure state}: For any pure state, we have the following Schmidt decomposition as 
\begin{align}
|\Psi\rangle = \sum^n_{i=1} \sqrt{s_i} |i\rangle_a |i\rangle_b,
\end{align}
with $\{ |i\rangle_a\} $ and $\{ |i\rangle_b\}$ are the orthonormal basis of the subsystems $a$ and $b$ respectively. We have, 
\begin{eqnarray}
C^{\text{min}}_{\mathcal{A}}(\rho^{ab}|\Pi)= 1-\sum_i s^2_i\geq 1-\frac{1}{n}, \nonumber \\
Q_{\mathcal{A}}(\rho^{ab})= \frac{\left( \sum_i \sqrt{s_i}\right)^2-1}{n}. ~~~~~~~~~~~~~~~ \nonumber 
\end{eqnarray}

\section{Application}
\label{applications}
In this section, we demonstrate the operational meaning of affinity-based coherence measure. Girolami et. al. identified a quantum correlation measure using Quantum Fisher Information (QFI), which determines the interferometric power of a quantum state. This quantity is more useful in the parameter estimation in a worst-case scenario and characterizes the nonclassical correlation of quantum system. This interferrometric power is defined as \cite{Girolami2014} 
\begin{align}
IP(\rho^{ab})=~^\text{~min~}_{|k\rangle \langle k|\otimes\mathds{1}} F(\rho^{ab}, H\otimes\mathds{1} )
\end{align}
by considering an interferometric setup and minimizing the quantum Fisher information over all the possible generators of a phase rotation on one party. Here $F(.)$ is quantum Fisher information defined via symmetric logarithm derivatives  and the minimum is intended over all fixed observables $H$ with non-degenerate spectrum $\{ \mu_i\} $. Considering the spectral decomposition of bipartite quantum state $\rho^{ab}=\sum_k \lambda_k|\psi_k\rangle \langle \psi_k|$  with $\lambda_{kl}=\frac{(\lambda_k-\lambda_l)^2}{(\lambda_k+\lambda_l)}$, we have 
\begin{eqnarray}
IP(\rho^{ab})=~^\text{min}_{U^a} F \left(\rho^{ab}, \sum_i U^a|i\rangle \langle i|U^a \otimes \mathds{1}\right)  = ~^\text{min}_{U^a} \sum_{k\leq l}  \lambda_{kl} |\langle \psi_k|\sum_i \mu_i U^a|i\rangle \langle i|U^{a\dagger}\otimes \mathds{1}|\psi_k\rangle  |^2 \nonumber \\
\leq ~^\text{min}_{U^a} \sum_{k\leq l}  \lambda_{kl} \sum_i |\mu_i||\langle \psi_k| U^a|i\rangle \langle i|U^{a\dagger}\otimes \mathds{1}|\psi_k\rangle  |^2  \nonumber \\
=~^\text{min}_{U^a} \sum_i |\mu_i| F(\rho^{ab},U^a|i\rangle \langle i|U^{a\dagger}\otimes \mathds{1}). \nonumber
\end{eqnarray}
Further, QFI is bounded  as \cite{SLuo20031}
\begin{align}
F(\rho^{ab},U^a|i\rangle \langle i|U^{a\dagger}\otimes \mathds{1})\leq 2I_{\text{WY}}(\rho^{ab},U^a|i\rangle \langle i|U^{a\dagger}\otimes \mathds{1}), \nonumber
\end{align}
where $I_{\text{WY}}(\rho^{ab})=-\text{Tr}[\sqrt{\rho^{ab}}, H\otimes\mathds{1}]^2$. Then the interferometric power is bounded by
\begin{align}
IP(\rho^{ab})  \leq  ~^\text{min}_{U^a} 2\sum_i |\mu_i|I_{\text{WY}}(\rho^{ab},U^a|i\rangle \langle i|U^{a\dagger}\otimes \mathds{1}). \nonumber
\end{align}
By choosing von Neumann projective measurement $\{ \Pi^a\}= \{U^a|i\rangle \langle i|U^{a\dagger} \}$, it is easy to show that $C_{\mathcal{A}}(\rho|\Pi)=I_{\text{WY}}(\rho^{ab},U^a|i\rangle \langle i|U^{a\dagger}\otimes \mathds{1})$. Hence, the interferometric power is bounded by
\begin{align}
IP(\rho^{ab})\leq C^{\text{min}}_{\mathcal{A}}(\rho|\Pi),
\end{align}
where the minimum is taken over all local unitary operations $U^a$. The affinity-based measure is thus interpreted as upper bound of interferometric power of a quantum state.

\section{Conclusions}
\label{Concl}
In this article, we have introduced a measure of coherence using affinity for quantum states. Having studied its properties, it is shown that this measure is a bonafide measure of quantum coherence. The affinity-based coherence measure is bounded by the coherence measures based on fidelity and trace distance. In addition, we have introduced a measure of bipartite quantum correlation based on affinity. Moreover, the trade-off relation between quantum coherence and the quantum correlation has also been established, showing that quantum correlation is limited by the coherence of marginal state under incoherent operations. Further, we propose a  measure of quantumness in terms of difference in bipartite coherence and its corresponding product state coherence  based on affinity. It is shown that this measure satisfies all the axioms of measure of quantum correlation. Finally, we have identified the operational meaning of affinity-based coherence as interferometric power of the quantum states.

\end{document}